*BAHI Camile, 2020*
*MPharm*

# 5-HT2A mediated plasticity as a target in major depression: a narrative review connecting the dots from neurobiology to cognition and psychology

# Abstract

As the world's primary morbidity factor, depression has a considerable impact on both an individual as well as a societal level. It is a complex and multifactorial mental health disorder, one whose pathophysiology is not yet fully understood. Despite their discovery several decades ago, classical antidepressants have been shown to provide limited benefits against this condition, yet no major enhancement since has been proven to treat major depressive disorder more efficiently.

However, substances such as ketamine and psychedelics (5-HT2AR agonists) have recently shown promising results and even received the grade of « Breakthrough Therapy » for this indication. The accurate mechanisms of action underlying the efficacy of these substances are still to be defined, but some similarities appear to be shared on different levels across these substances. These include their structural, functional and behavioral plasticity promoting abilities, as well as their capacity to promote Brain-Derived Neurotrophic Factor overexpression, which seems to constitute a key element underlying their immediate and long-lasting action.

From this observation, the present review aims to examine and connect the pharmacological pathways involved in these therapies to the neurobiological, cognitive and psychological responses that could be shared by both 5-HT2AR agonists and NMDA antagonists. It is suggested that BNDF overexpression resulting from mTOR activation would mediate both structural and functional plasticity, resulting in connectivity changes among high-level cognitive networks such as the Default Mode Network, finally leading to an increased and long-lasting psychological flexibility.

Connecting these pieces of evidence could provide interesting insights about their precise mechanisms of action and help researchers to develop biomarkers for antidepressant response in the future, contributing at the same time to getting a clearer idea concerning several components of MDD. If the hypotheses suggested in this review are verified by further trials, they could also constitute a starting point to developing safer and more efficient antidepressants, as well as provide information about the interactions that exist between different neurotransmitters systems.

# Abbreviation list

5-HIAA : 5HydroxyIndolAcétic acid
5-HT : Serotonin
AAQ-2 : Acceptance and Action Questionnaire
ACTH : AdrenoCorticoTrophic Hormone
ATU : temporary authorization of use
BDNF : Brain-Derived Neurotrophic Factor
BBB : blood brain barrier
BOLD : blood-oxygen-level dependent
CIA : Central Intelligence Agency
CIM-10 : international classification of diseases
CRH : Corticotrophin Releasing Hormone
CRP : C-reactive Protein
DAS : Death Anxiety Scale
DASS-21 : Depression, Anxiety and Stress Scales
DHF : dihydroxyflavone
DMN : Default Mode Network
DMT : Dimethyltryptamine
DMSO : diméthylsulfoxyde
DOI : 2,5-Dimethoxy-4-iodoamphetamine
Dr : Doctor
DSM : Diagnostic and Statistical Manual of Mental Disorders
DTS : Death Transcendance Scale
ECT : Electroconvulsive Therapy
FDA : Food and Drug Administration
GRID-HAM-D/A : GRID-Hamilton Depression/Anxiety Rating Scale
GC : GlucoCorticoides
HADS : Hospital Anxiety and Depression Scale
IL : InterLeukin
MAOI : Monoamine oxidase inhibitors
MDD : major depresive disorder
MRI : Magnetic Resonnance Imagery
fMRI : Functionnal-MRI
RS-fMRI : FMRI during resting state
CSF : cerebrospinal fluid
LSD : Lysergic Acid Diethylamide
LTP : Long Term Potentiation
MEQ30 : Mystical Experience Questionnaire
mPFC : Medial Prefrontal Cortex
mTOR : mechanistic Target Of Rapamycin
NMDA : N-méthyl-D-aspartate
WHO : World Health Organisation
PCC : Posterior Cingulate Cortex
PFC : PreFrontal Cortex
PIQ : Psychological Insight Questionnaire
CNS : Central Nervous system
STAI : State Trait Inventory Anxiety
tDCS : Transcranial Direct Current Stimulation
TrkB : Tyrosine Receptor Kinase B
TNF alpha : Tumor Necrosis Factor alpha

# Introduction: state of the art of major depressive disorder and opportunities opened by neuroplasticity

According to the WHO, depression represents the first morbidity factor in the world and would affect 300 million people (OMS, 2008).

This mood disorder is included in the Diagnostic and Statistical Manual of Mental Disorders (DSM-5) and the International Classification of Diseases (ICD-10) (OMS, 2008).

The consequences of this disease are serious and can become dramatic, leading to weakened social relationships and incapacitated professional life.
Depression is also correlated to metabolic, cardiovascular and infectious diseases, and represents the first suicide risk factor. Suicide rates increase 21-fold for patients with major depressive disorder (Olié et Mouren, 2014).
Classic antidepressant treatments such as tricyclic antidepressants, monoamine oxidase inhibitors (MAOI), and serotonin recapture inhibitors have all been developed on the basis of the monoaminergic hypothesis of depression, first introduced in the early 70s. It hypothesizes that an abnormally low monoaminergic activity in the brain would underlie a major depressive episode and that the above-mentioned anti-depressants would be effective in balancing that deficit.
This theory relies on the argument that a reduced amount of serotonin metabolites has been observed in the cerebrospinal fluid (CSF) of depressed patients without antidepressant treatment by Ashcroft and his team in the 50's.
Reduced concentrations of Serotonin (5-HT) and 5HydroxyIndolAcétic acid (5-HIAA) were also observed postmortem in depressed and suicide victims' brains. Additionally, patients with MDD presented lowered plasmatic concentrations of tryptophan, a serotonin precursor (Owens and Nemeroff, 1994).
Unfortunately, this hypothesis alone cannot be taken as a model of depression, and the large STAR*D study, conducted in 2014 shown that fifty percent of depressive patients are considered to be treatment-resistant.

Depression is nowadays considered to be a complex, multifactorial disease that must be modelized in the light of multiple complementary and imbricated hypotheses: Firstly, depression seems to have a strong hereditary component, according to studies conducted in twins, which show a significantly stronger prevalence of this disease in individuals that share the same genetic heritage (Su et al., 2009, Sullivan *et al.,* 2000).

Among the theories aiming to explain this pathology, one of the most studied recently is the inflammatory hypothesis of depression, based on the neuroinflammation phenomenon. Neuroinflammation depends on to two major mechanisms: a central mechanism that involves neuronal and astroglial production of cytokines (Fleshner et al., 2017; Wohleb et al., 2016), and a peripheral mechanism, which enables cytokines to reach the CNS either through active transport or passive diffusion from one side to another of the BBB (blood brain barrier) (Quan and Banks 2007).

Peripheral cytokines can also act on the CNS via the vagus nerve (CN X) activation – involving intestinal microbiota– and through the entry of activated peripheral monocytes in the CNS (Krishnadas and Cavanagh 2012).

The inflammatory hypothesis of depression is reinforced by several arguments: first, the observable co-morbidity between depression and diseases related to the immune system (especially cardiovascular and auto-immune diseases; Halaris 2017, Pryce et al., 2017), but also the observable modulation of cytokines rates in depressive disorder, such as increased rates of TNF alpha, IL-6 and CRP (Huang et al., 2018; Zhang et al., 2018; Zou et al., 2018; Danner et al., 2003; Dowlati et al., 2010). These increases seem to be correlated to the severity of the depressive episode (Zou et al., 2018).
Significantly higher rates of several pro-inflammatory factors, such as TNF-alpha and IL-2 have been reported in patients that have committed a suicide attempt (Janelidze et al., 2011, Pandey et al., 2012).
In addition, this theory is supported by the fact that administrating exogenous cytokines (for example, giving interferon to patients with cancer) can induce depressive symptoms (Krishnadas and Cavanagh 2012, Capuron and Miller 2011), and that anti-inflammatory treatments and anti-cytokines can have a positive impact on mood (Liu et al., 2017; Raison 2017; Köhler et al., 2014; Kappelmann et al., 2016).

The implication of intestinal microbiota in depression has also been a major point of interest recently.
A microbiota is represented by the commensal (nonpathogenic) flora, constituted of microorganisms in a specific environment. More specifically, intestinal microbiota represents the largest human microbiota, weighing around 2kg, containing 2 to 10-fold more cells than the total number of cells constituting the human body.
This microbiota has a role in digestion, ensures proper functioning of the epithelium, and modulates the intestinal immune system (Doré et al., 2017).
According to Bailey, 2007, the intestinal microbiota could be implicated in MDD when modification of the homeostatic interaction between the microbiota and its host occurs, under the influence of stress –especially by means of glucocorticoids–. This phenomenon generates an increase in microbiota's activity, causing oxidative reactions, and resulting in an increase in inflammation and therefore in increased cytokines productions.
Several studies have shown that the rates of some bacterial species constituting the microbiota are significantly altered in patients with MDD when compared to controls (Lin et al., 2017; Aizawa et al., 2016). Accordingly, the administration of probiotics might significantly decrease depressive symptoms, according to a meta-analysis by Akkasheh et al. conducted in 2016.

Lastly, the hypothalamic-pituitary-adrenal (HPA) axis hypothesis relies on the pre-existing link between chronic stress and depression. This theory takes place at a central position in the pathophysiology of depressive disorder, interacting directly with the inflammatory and the microbiota arguments, but also with the monoaminergic hypothesis.

Under normal conditions, stress leads to CRH (Corticotrophin Releasing Hormone) and vasopressin release from the hypothalamus, leading to ACTH (adrenocorticotrophic hormone) release from the pituitary gland, finally leading to glucocorticoid (especially cortisol) production by the adrenal gland.
Glucocorticoids act both at the peripheral and the central level, causing a disturbance in neuronal survival, neurogenesis, memorization, and emotional appreciation (Herbert et al., 2006; Pariante and Lightman 2008; Nicolaides et al., 2015).
In standard conditions, this system is subject to feedback: a large amount of GC causes a decrease in CRH and vasopressin release, ensuring homeostasis. In depression, this feedback system would be dysfunctional due to chronic stress exposure.
In this situation, there is a decrease in monoamine levels (including serotonin) due to a redirection of tryptophan metabolism towards kynurenines' production (under the action of cortisol activating Nf-kB factor) also leading to an increase in cytokines levels (Miller et al., 2009).
To support this hypothesis, it is important to note that abnormally high cortisol levels are found in saliva, plasma and urine of depressed patients, although these results are inconsistent (Nemeroff et Vale 2005).

Furthermore, neurotrophic factors involved in neuronal and synaptic plasticity such as BDNF (Brain-Derived Neurotrophic Factor), have a preponderant role within this last hypothesis:

It has been demonstrated that stress induces a decrease in the expression of BDNF; for example, in the dentate gyrus and hippocampus more generally, in animals (Smith 1996) and that a decrease in plasmatic BDNF, correlated to central concentration, is observed in depressed patients. Conversely, the majority of classic antidepressants induce an increase in BDNF concentrations, making it possible to reverse the effects of stress involved in depression (Nibuya et al., 1996).
Moreover, overexpression of cortical BDNF has also been observed in patients that were responsive to antidepressant treatments (Shimizu et al., 2003).

These elements suggest that there is a reachable therapeutic path involving BDNF and the neuronal and synaptic plasticity it promotes.

According to this perspective, it is useful to mention that innovative antidepressant treatments, involving various mechanisms of action, share the capacity to induce BDNF overexpression and promote neuronal plasticity, which seems to be a decisive and necessary factor for their long-term efficacy:

Ketamine, a molecule with anesthetic and dissociative proprieties, whose action is lying principally on glutamate NMDA-R antagonism, has recently received FDA approval for its use against depression and also has a Temporary Authorization of Use in France and other European countries for the same application (FDA, 2019).
Its long-term efficacy seems to rely on the BDNF overexpression that it provokes through modulation of glutamatergic system and on its ability to promote neuronal plasticity, which results from this overexpression (Woelfer et al., 2019). However, the exact mechanisms underlying the efficacy of this molecule remain unclear (Yang et al., 2019).

Psychedelics, a pharmacological class defined by their 5-HT2A receptor agonist properties and including hallucinogenic molecules such as lysergic acid diethylamide (LSD), psilocybin, N,N-dimethyltryptamine (DMT) or mescaline, are also studied for their application in depression. Psilocybin recently received the grade of "breakthrough therapy" from the FDA in this indication. These substances reveal a promising efficacy in depression, including treatment-resistant forms. (Griffiths, 2016; Ross, 2016; Carhart-Harris et al., 2016; see Bahi, 2018; Reiche et al., 2018 for reviews).
These molecules are of particular interest for several reasons: first of all, they are effective against depression in an immediate and lasting way over more than 6 months after a single dose, they also cause a BDNF overexpression (which seems to be a key element of their efficacy) and have a strong specificity of action on the 5HT2A receptors, a serotonin receptor subtype located mainly on layer V pyramidal glutamatergic neurons (Carhart-Harris and Nutt, 2017, Vollenweider and Kometer, 2010). Moreover, their therapeutic and subjective effects are completely blocked by administrating an antagonist of these specific receptors (Preller et al., 2019; Pitts et al., 2017).
In addition, although some 5-HT2A antagonists (such as mirtazapine) have a positive outcome on the symptoms of depression, no pure 5-HT2A-R antagonist has shown any efficacy in this pathology (Ebdrup et al., 2011).
This suggests that the efficacy of these antagonists is dependent on other associated pharmacological mechanisms –blocking 5HT2A-R could facilitate agonism on 5HT1A postsynaptic receptor– and their sedative effect that contributes to reduce stress (Idzikowski et al., 1987; Teegarden et al., 2008; Vanover and Davis, 2010).

These specificities place 5HT2A-R agonists in an interesting position for studying interactions between the serotoninergic system (involved in the efficacy of most anti-depressive therapies) and other neurotransmitters systems such as glutamatergic system, whose involvement in neuroplasticity through the BDNF is now well established (Cohen-Cory et al., 2010).

An interesting model that has been developed recently implies that an increase in structural neural plasticity accompanied by changes in functional connectivity could be correlated to an increase in psychological flexibility in depressed patients. Given that MDD is characterized by mental rigidity and a pessimistic vision of the self and the world, restoring the capacity for action and active coping in patients would play a decisive role in the efficacy of these substances against depression (Carhart-Harris and Nutt, 2017; Davis et al., 2020).

Studying the link between the systems exposed in this model could not only provide tangible arguments in favor of this model but also give the possibility to develop specific therapies targeted on the identified neurobiological mechanisms. It could also open new ways to develop reliable biomarkers to monitor the response to these therapies, which is currently lacking in MDD and generally in psychiatry.

Accordingly, this review will focus on psychedelics by exploring their mechanism of action, aiming to connect pieces of evidence that support the exposed model both at a psychopharmacological, cognitive and psychological level.

# Psychedelics in depression: an effective therapeutic class and a tool for understanding plasticity

## A brief history of psychedelics

Psychedelics *stricto sensu*, characterized by their 5-HT2AR agonist properties and their hallucinogenic effect, are present in natural sources and used since immemorial times by different cultures around the world.

Indeed, vestiges dedicated to the "mushroom god" have been found in Central America, dating from the year 1000 to 500 BC, a statue of a deity holding "sacred" plants in his hands, including Psilocybin mushrooms, dating from the 16th century was also found in Mexico and other vestiges related to Psychedelics were discovered around the world (Schultes, Hoffman, 1992).

It is only later, in 1943, following the LSD discovery by the chemist and researcher Albert Hofmann for Sandoz pharmaceuticals, that psychedelic properties began to receive interest from the scientific community. Later, in the 1950's, Richard Evans Shultes returns from Central America, bringing Psilocybin mushrooms specimens to the Harvard herbarium.
Hofmann then isolates the chemical compound called psilocybin in 1958 and synthesizes it in 1959. Sandoz Laboratories decides to submit these two interesting substances to clinical research and start to produce psilocybin in 1960. Their pill formulation contains 2mg of pure psilocybin and is sent to voluntary psychiatrists and psychologists (Erowid, 2019).

A poorly supervised research phase was begun with various applications, ranging from self-administration by psychiatrists in order to familiarize themselves with a possible model of psychosis, to the conducting of psychedelic-assisted psychotherapy sessions, with the aim to solve psychological problems such as depression.

Unfortunately, the lack of a rigorous scientific framework in these researches, the misuse of these substances by the "beat generation", as well as unsuccessful experiments conducted by the CIA for making a military use out of them, caused the cessation and prohibition of psychedelic research around 1977 all over the world (Erowid, 2019).

It has been necessary to wait until the end of the 90s for research on psychedelics to start again, though timidly (Erowid, 2019) and much longer –until around 2016– to see the foundation of research centers specifically dedicated to psychedelic research. some of these centers are: The Heffter Research Institute, Beckley's foundation, the Psychedelic Research Center of the prestigious Johns Hopkins University in Baltimore, founded in September 2019, that of Imperial College of London at the same period, but also the companies Compass Pathways and Usona institute (which are supported by major global health funds and empowered to conduct multicentric phase 2b clinical trials) or the European Foundation for Psychedelic Science: MIND foundation, based in Berlin and founded in 2016, which is about to conduct the second largest study in the world to assess the efficacy of psilocybin against depression this year.

## Contemporary research: proof of concept, clinical efficiency, applications

The enthusiasm expressed by these pioneers for this class of molecules has ultimately appeared to be justified: recent research on psilocybin demonstrates clinical interest in the treatment of depression, addictions, and anxiety. Psychedelics could also help in the more general aim of bringing a better understanding of cognitive functions and several pathologies, such as bipolar disorders or schizophrenia.

In fact, despite the recent revival of psychedelic research, promising examples already exist: Ross and colleagues conducted a trial in 2016 to assess the efficacy of psilocybin in 29 patients with life-threatening, cancer-related depression and anxiety.

This randomized, placebo-controlled, double-blind study, included cancer patients who met DSM-IV criteria for anxiety or depression were administered either psilocybin at a dose of 0.3 mg/kg or niacin at a dose of 250mg (the latter being administered as an active control, for the physical rush it provokes).
A cross-over between the two sessions took place seven weeks after the first session.

The primary outcome of this study was to measure anxiety and depression before crossover. Secondary outcomes measured before and after the crossover included: depression score evolution, the measurement of existential distress, quality of life and spirituality, but also the measurement of the immediate and persistent effects of psilocybin on subjective experience, cognition, affects and behavior.

Depression and anxiety levels were assessed using internationally recognized rating scales such as HADS (hospital anxiety and depression scale), STAI (Spielberger straight trait anxiety inventory) or BDI (Beck depression's inventory).

Results after the first session for the "psilocybin first" group of patients showed an immediate and significant response for depression and anxiety.
These results were found to remain significant at all follow-up points, including six months after the active session.

The group receiving Niacin before the crossover, as expected, did not show any significant improvement whether it was durable or not.
However, after the second session –the session distributing psilocybin for the patients in this treatment group–, as for the "psilocybin first" group, significant response rates were observed for depression and anxiety, and remained present even six months after the dosing session.

About the safety of use, no serious adverse events, either psychiatric or medical, have been reported regarding Psilocybin.
No psychiatric intervention (such as benzodiazepines administration) was necessary and no patient abused or became addicted to psilocybin (Ross et al., 2016).
Only minor, tolerable and transient anxiety states have been reported and are not necessarily considered to be a negative outcome, in the sense that they would, in fact, be part of the patient's path towards their personal problems and would therefore be linked with treatment efficacy (Swift et al., 2017 ;  Belser et al., 2017).

Concerning biological monitoring (blood pressure and heart rate) there was an increase in both these markers, but these were transient and not serious, and correspond to known effects of psilocybin (Ross et al., 2016).

These results are in line with the study conducted by Krebs and Johansen in 2013, based on data from the national survey on drug use and Health from 2001 to 2004, which found no association between the use of psychedelics and any increased rate of mental illness (psychosis, schizophrenia, HPPD, etc.). "Magic mushrooms" have also been shown to be the safest drug for recreational use in the 2017 Global Health Survey, the largest study realized on the subject (Winstock et al., 2017).

Another randomized double-blind study, considered to be the most rigorous to date about this topic, was carried out by Griffiths and colleagues in 2016 on 51 patients, with the same primary outcome. This time, it compared a therapeutic dose of psilocybin (22 or 30 mg/70kg), to a dose considered to be inactive and used as a placebo (1 or 3 mg/70kg).
The study design was similar to the study previously presented in this review, with two crossover sessions, and used the same scales to assess depression and anxiety, with the additional use of the GRID-HAM-D (GRID-Hamilton Depression) scale for depression and HAM-A  (GRID-Hamilton Anxiety) for anxiety. The results of this study were found to confirm results observed in the trial realized by Ross and his team.
Indeed, no medical intervention was necessary on the physical or psychiatric level during or following the sessions and no adverse events or addiction were to be deplored. Only transient anxiety was reported, as well as headache, during or after the high-dose session, in few participants (Griffiths et al., 2018).
From an efficacy point of view, again, the group that received a high dose of Psilocybin first showed a 92% response rate (32% in the "low dose first" group) and these results were still observable after 6 months in 80% of the patients in this group.
On the HAM-A scale, the results were 76% for response rate (versus 24% in the control group) and 83% at 6 months. After merging the results of the two sessions, response rates for depression and Anxiety were 78 and 83% respectively and remission rates were 65% and 57%, respectively (Griffiths et al., 2018).

The results of these two studies are very encouraging. Thereby, they were the subject of a meta-analysis that also presented two further qualitative studies on patients' experience with psilocybin. The meta-analysis aimed to ensure that the results obtained were statistically significant when compared to placebo and when combining the results of the two studies, thanks to the use of a Peto ODDS ratio (Figure 1) (Bahi, 2018).
The results obtained with the psilocybin group were found to be significantly and largely superior to the results obtained for the placebo group.

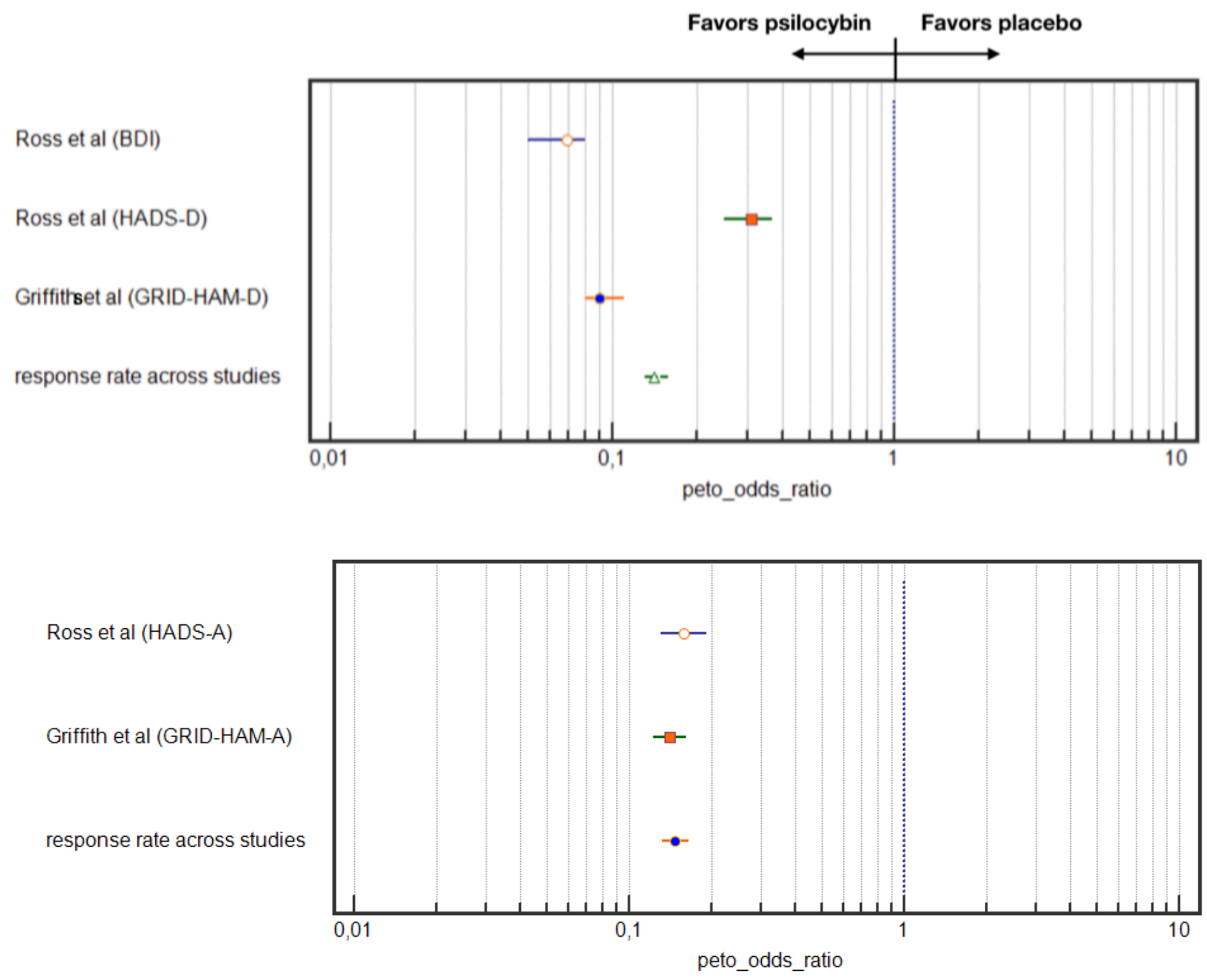

**Figure 1. Psychedelics show a significantly better response rates on depression and anxiety compared to placebo.**
(Bahi, 2018)

Phase 2b clinical trials are currently underway intending to use psilocybin in treatment-resistant depression. The companies Compass Pathways and Usona institute have received the “breakthrough therapy” status from the FDA for studies on this application. The MIND foundation is expected to start the second largest study in the world to assess the efficacy of psilocybin against depression around summer 2020.
Other fields of application are envisaged, like the treatment of addictions; in this line, Johns Hopkins University is launching a clinical trial with the aim of treating nicotine-addicted patients in 2020 (Johns Hopkins University, 2019).

s

## Mechanism of action

The efficacy of these potential treatments in depression is therefore firmly assessed, and they are on their way to obtaining a market authorization in the upcoming years; however, the mechanisms underlying their effect have not yet been fully elucidated.
Research on this topic shows an increase in neuronal plasticity probably involving an increase in the level of neurotrophic factors such as BDNF, as well as modification of the connectivity in neural networks involved in resting-state activity, such as the DMN or default mode network, which is physiologically responsible for high-level cognition functions such as introspection and rumination (the latter being more specific to MDD) and is mainly implying an anteroposterior functional and blood coupling between the PCC (posterior cingulate cortex), the mPFC (medial prefrontal cortex) and the amygdala (Carhart-Harris et al., 2016).

Several articles highlight the fact that these changes in structural plasticity and functional connectivity within networks could be linked and superposed to the behavioral improvements associated with treatment response (Carhart-Harris and Nutt, 2017; Ly et al, 2018). This review will focus on exploring this link.
Furthermore, it is interesting to note that connectivity changes in these resting-state networks are implied in other therapies previously mentioned, such as ketamine, but also in tDCS (transcranial direct current stimulation) and electroconvulsive therapy (Woelfer et al., 2019; Keezer et al., 2011; Carhart-Harris and Nutt, 2017).
Other researchers highlight the specificity to which the action of psychedelics is subject through 5HT2A agonism, by demonstrating that the blockade of these receptors by a specific antagonist such as ketanserin, both abolishes the hallucinogenic effect in humans and the effect on in vitro neuronal plasticity mediated by these molecules (Ly et al., 2018; Preller et al., 2019).
Moreover, we now know that 5-HT2AR agonist action positively modulates glutamatergic neurotransmission (figure 2) and activates signaling pathways involving B-arrestine, for example (Vollenweider and Kometer, 2010), a protein engaged in neuronal plasticity pathways such as the mTOR signaling pathway (Jean-Charles et al., 2017).

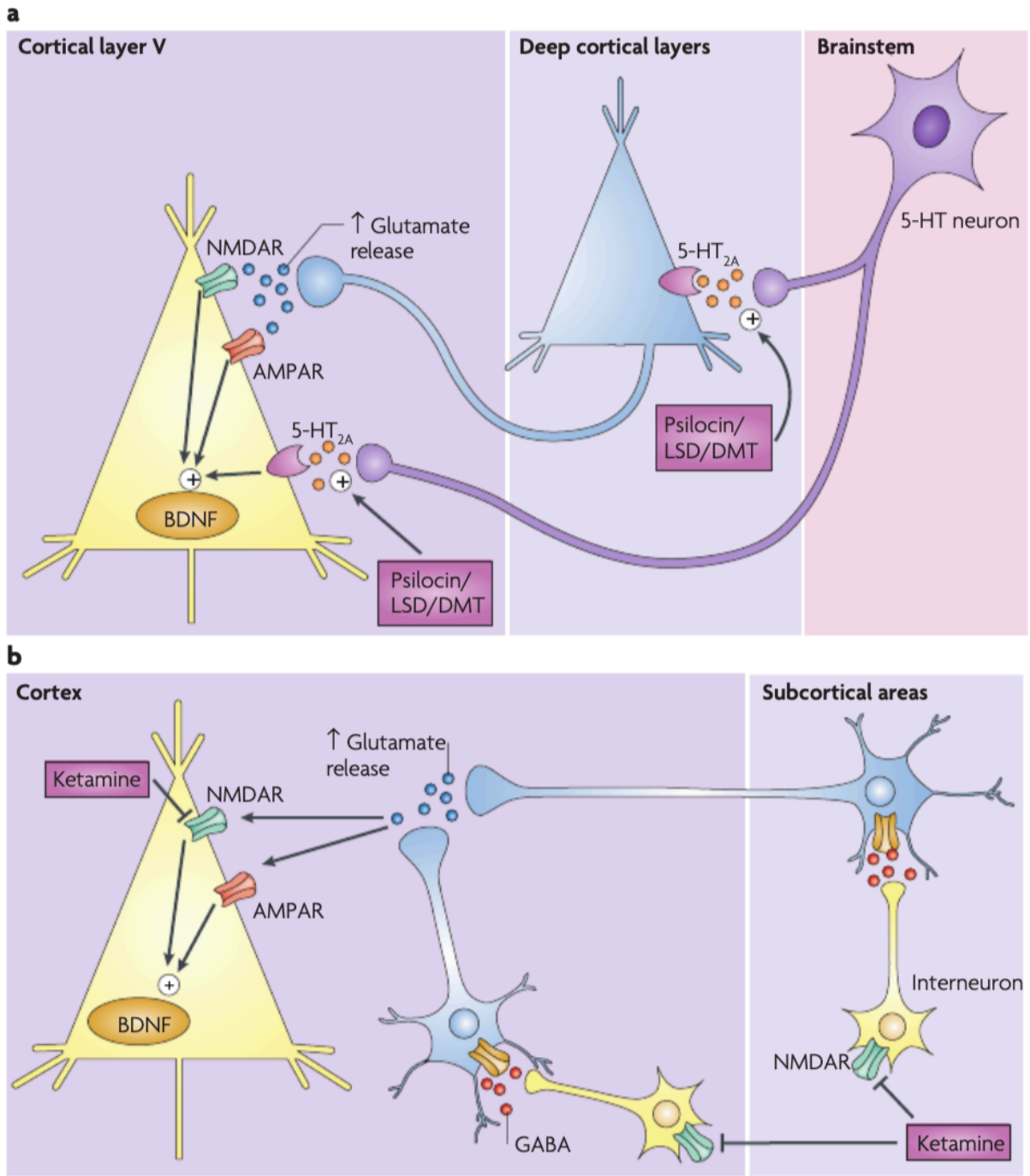

**Figure 2. Activation of the prefrontal network and glutamate release by psychedelics** (Vollenweider and Kometer, 2010)

# Unraveling implied mechanisms: from molecular targets to network connectivity.

## A common signaling pathway shared by psychoplastogens, also involved in functional connectivity changes?

### Exploring signaling pathways

As discussed above, one of the key mechanisms involved in the effect of psychedelics depends on their 5HT2A-R agonism properties, causing a hallucinogenic effect, which is involved in their efficacy against depression (swift et al., 2017; Davis et al ., 2020; see Bahi, 2018 for review).
This property would also underlie the overexpression of neurotrophic factors such as BDNF, as well as changes in the activity and connectivity within several resting-state functional networks (i.e. engaged when no specific task is proposed during functional MRI).
These two last points seem to be necessary for psychedelics' immediate and long-term efficacy.
These arguments are reinforced by the fact that altered anteroposterior connectivity in specific DMN regions is often observed in MDD and is associated with the ruminative aspect of MDD which represents an important component of this syndrome (Carhart-Harris and Nutt 2017).

To verify that the neuroplasticity occurring following the use of psychedelics is linked with their 5-HT2AR agonist properties and to get a closer look at the signaling pathways that underlie this mechanism, Ly's team carried out a series of experiments based on results observed with ketamine.
Results confirm the initial hypothesis: the psychedelic molecules tested were able to increase arbor complexity in rats' cortical neurons, showing an increase in the number of dendrites' branches and also an increase in the total length of the arbor, without however causing an increase in the number of primary dendrites, nor the size of the longest dendrites.

7,8- dihydroxyflavone (DHF) was used as a positive control because of its well-known plasticity-promoting properties and its different structure and molecular target compared to psychedelics. This control is also able to promote complexification of the arbor, while negative controls such as Serotonin or D-Amphetamine are not.

Then, Ly's team determined the relative potency in plasticity induction using an 8 points dose-response study with an effect varying from 0 to 100%, with 0% being defined by the effect of the vehicle (DMSO, which is a solvent) at a concentration of 0.1%, and 100% being defined by a 10umol Ketamine concentration (corresponding to the maximum brain concentration obtained after an intraperitoneal injection of Ketamine in rats).

Surprisingly, psychedelics constantly showed better results than ketamine at the same dose.
These results were verified in vivo, using Drosophila Larvae as an experimental model and LSD (ergot-derived psychedelic molecule) and DOI (which has an amphetaminic structure) as the active molecules (Ly et al., 2018).

Again, the number of dendritic branching in class 1 sensory neurons was found to be increased, without however observing an increase in the total length of dendritic arbors.
These results, confirmed in both vertebrates and invertebrates, suggest that the mechanism underlying this plasticity promoting abilities is evolutionarily conserved (Ly et al., 2018).

Dr. Ly and colleagues then wanted to assess the effects of psychedelics on synaptogenesis and spinogenesis in rats. They proceeded in vitro first, by exposing rat cortical neurons to DOI, DMT (tryptamine psychedelic) and LSD, and in vivo afterward, with a 10mg/kg DMT peritoneal injection. Results demonstrated an increase in the synapse density and an increase in the number of dendritic spines (up to 2 times more spines), without however showing any change in the dendrite length in vitro (figure 3).
In vivo, these results were mainly observable on pyramidal cortical neurons (which is also the preferred localization for 5HT2A receptors in the cortex) and with an intensity comparable to standards results observed with Ketamine at the same dosage (Ly et al., 2018).

It is important to note that these remarkable effects on structural neuroplasticity were accompanied with changes at the functional level in the PFC: similar increases in the amplitude and frequency of spontaneous excitatory postsynaptic currents (EPSCs) were observed at both 10mg/kg and 1mg/kg concentrations (hallucinogenic and sub-hallucinogenic doses).
Furthermore, the fact that the half-life of DMT is only 15 minutes and that the functional changes were still observable hours after this delay, seems to indicate that the plasticity promoting abilities of psychedelics can last after total clearance of the substance from the body (Ly et al., 2018).

These changes also mirror those produced by Ketamine (Li et al., 2010). This underlines an additional parallel between psychedelics and Ketamine: indeed, these two classes of psychoplastogens seem able to cause similar plasticity changes both at the structural and functional levels in an immediate and lasting way. These two types of substances are also able to facilitate fear extinction learning, a behavioral mechanism involved in stress response (Cameron et al., 2018; Girgenti et al., 2017), in which the PFC is predominantly involved (Quirk et al., 2006).
In addition, fear extinction learning can be improved by increasing BDNF concentration in the PFC and the behavioral effects of Ketamine have been shown to be BDNF-dependent (Lepack et al., 2014).
The role of neurotrophic factors in neuroplasticity is well established today (Cohen-Cory et al., 2010) and various studies have shown that psychedelics may be able to increase the levels of neurotrophic factors, especially BDNF (He and al., 2005; Martin et al., 2014; Nichols and Sanders-bush, 2002, Vaidya et al., 1997).

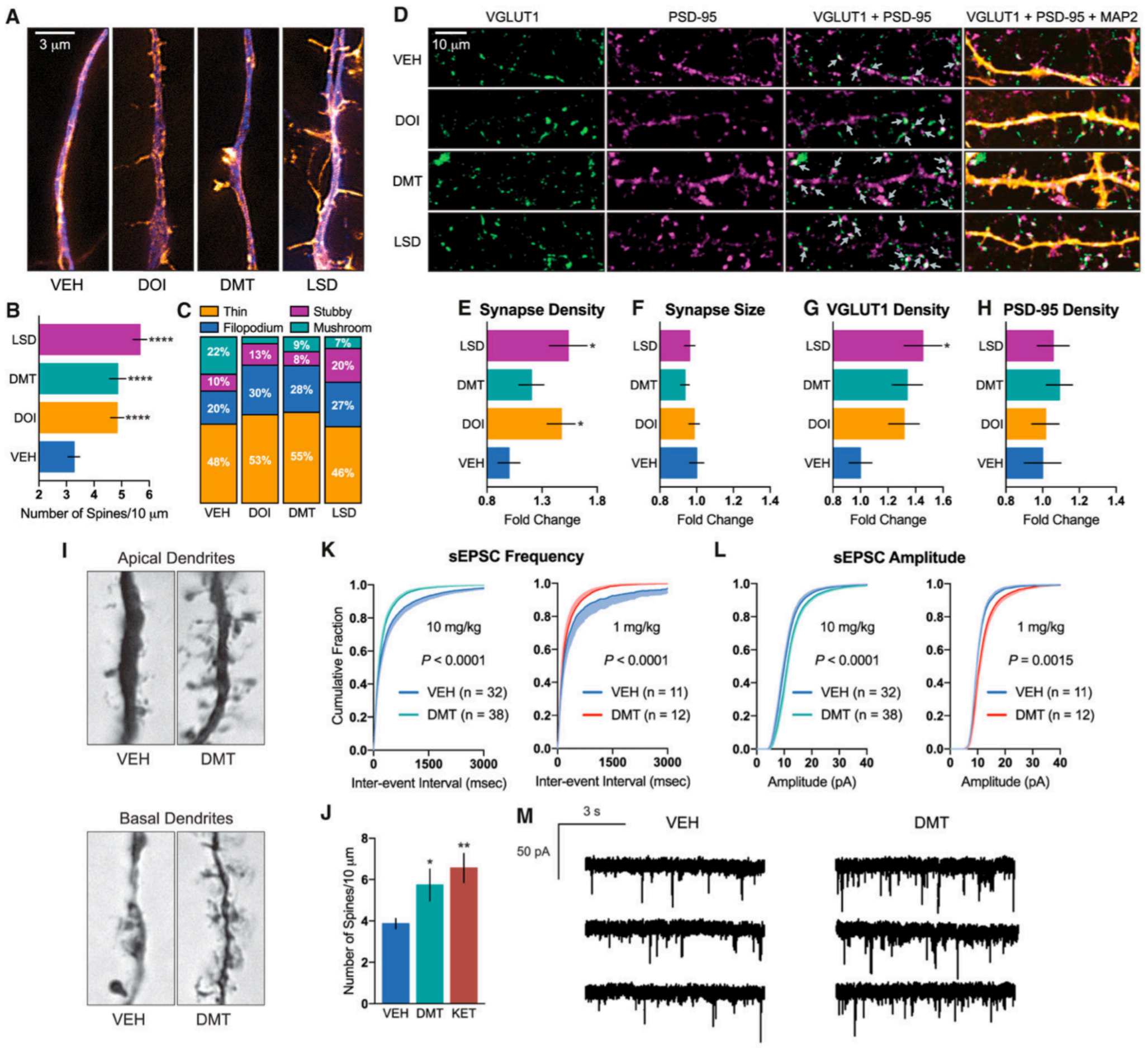

**Figure 3. Psychedelics Promote Spinogenesis, Synaptogenesis, and Functional Plasticity**

(A) Representative images of cortical neurons (DIV19) treated with compounds for 24h, demonstrating that psychedelics increase the number of dendritic spines (blue, MAP2; orange, F-actin).

(B) Quantification of spine density (n = 56–65 neurons).

(C) Relative proportions of spine types following treatment of cortical cultures with psychedelics (n = 16–21 neurons).

(D) Representative images of cortical neurons (DIV19) treated for 24 hr, demonstrating that psychedelics increase synaptogenesis (green, VGLUT1; magenta, PSD-95; yellow, MAP2). White areas in the VGLUT1 + PSD-95 images indicate colocalization of pre- and postsynaptic makers and are indicated by gray arrows.

(E–H) Quantification of synapse density (E), synapse size (F), presynaptic density (VGLUT1) (G), and postsynaptic density (PSD-95) (H) following 24-hr treatment of cortical neurons (DIV19) (n = 39–42 neurons).

(I) Representative images of Golgi-Cox-stained pyramidal neurons from the PFC of rats 24 hr after receiving a 10 mg/kg dose of DMT.

(J) Quantification of spines from (I), demonstrating that DMT (10 mg/kg) increases spinogenesis in vivo to a comparable extent as ketamine (10 mg/kg) (n = 11–17 neurons).

(K and L) Whole-cell voltage-clamp recordings of layer V pyramidal neurons from slices obtained 24 hr after DMT treatment (10 mg/kg and 1 mg/kg) demonstrate that DMT increases both spontaneous excitatory postsynaptic current (sEPSC) frequency (K) and amplitude (L) (n = 11–38 neurons from 3 animals).

(M) Representative traces for the 10 mg/kg experiments quantified in (K) and (L).

*$p < 0.05$, **$p < 0.01$, ***$p < 0.001$, ****$p < 0.0001$, as compared to vehicle control (VEH). Data are represented as mean ± SEM. See also Figure S6.

(Ly et al., 2019)

It is therefore natural that Ly's team sought to determine the link between BDNF and psychedelics' ability to promote structural and functional plasticity.
To verify it, they first treated rat cortical neurons with BDNF, DOI (10um) and a combination of the two, in order to explore if there was a possible synergistic or additive effect: 50 mg/mL BDNF increased spinogenesis and synaptogenesis to a comparable extent as 10uM DOI. Moreover, a combination of the two substances did not confer any benefits to neuroplasticity (Ly et al., 2018). This implies that the mechanism promoting neuroplasticity mediated by these two substances is in fact shared.

Cultured neurons were then treated with psychedelics (DOI, LSD, DMT) for 24 hours to monitor BDNF gene and protein expression.
No changes in gene transcripts were observed. However, BDNF protein levels were two times higher. This increase was not significant, but blockade of BDNF's high-affinity receptor Trkb with the high-affinity antagonist ANA-12 completely abolished any BDNF or psychedelics effects on synaptogenesis or spinogenesis (Ly et al., 2018).

Since Trkb receptor activation leads to mTOR signaling (which is essential for promoting neuronal plasticity and neurotrophic factors production) and that the response to Ketamine is dependent on this pathway, Ly and colleagues decided to treat cultured neurons with Rapamycin, a specific mTOR inhibitor, to find out whether it was involved in plasticity mediated by psychedelics.

Without surprise, psychedelic-induced neuroplasticity was completely blocked following Rapamycin exposure. These results confirm that mTOR pathway plays a decisive role in the plasticity promoted by psychedelics (Ly et al., 2018).
By specifying the signaling pathways involved in the plasticity promoting properties of psychedelics, these results bring further details to Vollenweider and Kometer's hypothesis formulated in 2010, suggesting a shared mechanism of action between Ketamine and psychedelics underpinning their efficacy in depression (Figure 4).

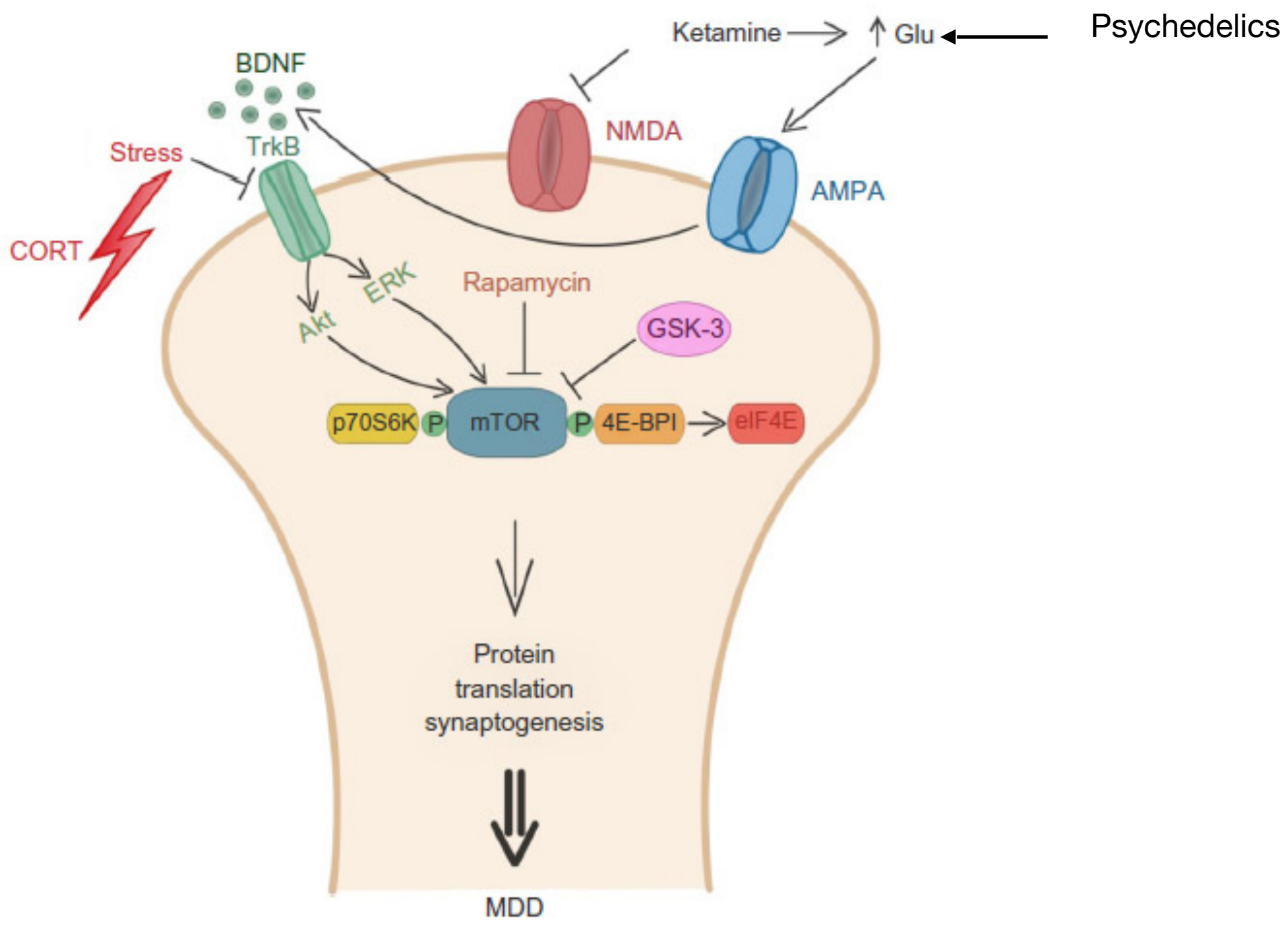

**Figure 4. Ketamine and Psychedelics action on depression through mTOR**
(According to Réus et al., 2015)

Lastly, Ly and colleagues aimed to verify that the plasticity-promoting proprieties of psychedelics was indeed caused by 5HT2A receptors' stimulation (since this stimulation is also responsible for the hallucinogenic effect, that their psychoplastogen proprieties are strongly correlated to their affinity for these receptors and that these receptors do not constitute a target for Ketamine, which appears to share the same signaling pathways).
The researchers verified first that cultured cortical rats' cells were indeed presenting 5HT2A receptors and then treated those cultures with psychedelics.
Then, they noted that blocking those receptors with a specific agonist: Ketanserin, at a 100-fold concentration compared with psychedelics (the concentration which induced the maximal effect in normal condition: 10Um) resulted in a complete blockade of the plasticity promoting proprieties of psychedelics tested (Ly et al., 2018).
It is, however, possible that Ketanserin blocked other receptors at this concentration. To limit this possible bias, they used lower doses of LSD (10nM) and progressively increased the associated Ketanserine dosage (Ly et al., 2018). At a dose of 10nM, Ketanserin was able to block 50% of LSD's psychoplastogenic effects for 10 nM, which is in line with the fact that these two molecules have a similar affinity for 5HT2A-R.
Accordingly, 100nM doses of Ketanserin were able to completely block the effect of 10nM LSD concentrations. At this dose, Ketanserin is a fairly specific 5HT2A-R antagonist (Ly et al., 2018).

These results prove the direct implication of psychedelics in plasticity promotion and give further details about their precise mechanism of action, mediated by 5HT2A-R and implicated in their efficacy against depression.
Ly's experiments also reveal that molecules with different structures (tryptamines, ergolines, amphetamines) and different molecular targets (Ketamine block NMDA-R while psychedelics stimulate 5HT2A-R) are sharing the same signaling pathways (involving BDNF, Trkb and mTOR) underlying their plasticity promoting properties. This is even more remarkable since Ketamine can lead to addiction, whereas psychedelics do not (Nutt et al., 2007, 2010) and that these effects are consistent both in vitro and in vivo, with species of different degrees of evolution.

Another notable element is that the activation of these neuroplasticity pathways seems to mediate the behavioral effects associated with the response to these substances through network connectivity changes. The amplitude of these changes seems to correlate with the psychoplastogenic potency of these substances (Ly et al, 2018).

This last point is particularly important for establishing the link between the molecular targets that promote neuroplasticity and associated functional connectivity changes in functional brain networks implied in depression (the focus point being on the  DMN, whose functional connectivity changes are involved in psychedelic and Ketamine antidepressant responses Carhart-Harris et al., 2016, Kraguljac et al., 2017; Woelfer et al., 2019).

Similar functional connectivity changes within the DMN are also observed using non-pharmacological techniques, such as ECT or tDCS, and seem to be correlated with the behavioral response to these therapies (Carhart-Harris and Nutt, 2017; Keeser et al., 2011). Nevertheless, the precise origin of these functional changes in resting state networks remains poorly understood. Specifying these pathways could then also be useful outside psychedelic therapy and even pharmacological therapies framework.

### Exploring mechanisms underlying functional connectivity changes.

Some research about Ketamine that has explored the origin of these resulting functional connectivity changes could also apply to psychedelics.
Researchers underline that the increase of peripheral BDNF levels correlates with functional connectivity changes observed after Ketamine infusion in a placebo-controlled, double-blind study carried out on 53 healthy volunteers by Woelfer and colleagues in 2019.

This study aimed to explore and assess this eventual correlation. They proceeded by collecting a blood sample in patients at three time points (before Ketamine infusion, 120 minutes and finally 24 hours after infusion), quantifying BDNF levels in these samples using the sandwich ELISA immunoenzimatic technique, and comparing this variation in plasma BDNF levels with the functional connectivity changes observed in brain areas that are known to show modified connectivity after Ketamine, using RS-fMRI (Woelfer et al, 2019).
Those regions include structures such as the mPFC and Posterior Cingulate Cortex, that superpose with the Default Mode Network and correspond to the structures where the functional connectivity is modified during psychedelic infusion (Carhart-Harris et al., 2016).

Statistical correlation analyses between these data were then carried out to verify the initial hypothesis: As expected, results showed a significant correlation between the variation in peripheral BDNF levels and changes in functional connectivity in the DMN for patients in whom Ketamine infusion led to an increase in BDNF levels. This correlation is particularly significant when looking at the disconnection between PCC and mPFC occurring during the effect of Ketamine (Woelfer et al, 2019).

On the other hand, the placebo group and patients whose BDNF levels were not increased following Ketamine infusion did not show any correlation between functional connectivity and variation in BDNF levels over 24 hours. These results support the hypothesis according to which a proportionality relation exists between BDNF overexpression and changes in functional connectivity, as well as the hypothesis according to which the response to Ketamine (and, supposedly, psychedelics) in depression might depend on the amplitude of BDNF overexpression in patients (figure 5) (Woelfer et al., 2019).

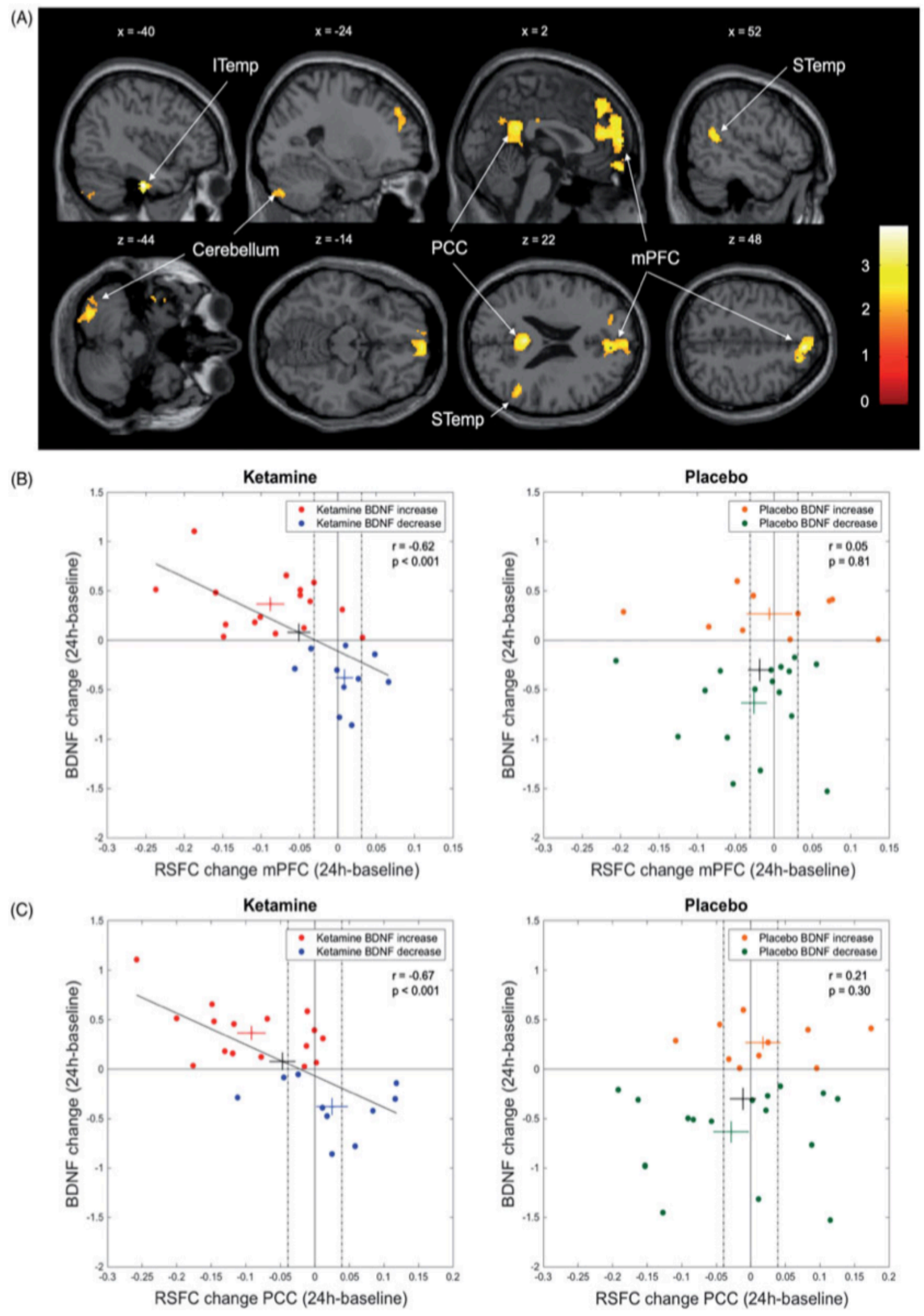

**Figure 5. Whole-brain regression analysis with interaction term to test the link between BDNF and resting-state functional connectivity (RSFC) changes at 24 h post-infusion.**

(A) Participants receiving ketamine or placebo showed a significantly differential association between changes in plasma BDNF levels after 24 h compared to baseline and dorsomedial prefrontal cortex (dmPFC) seeded RSFC changes to ventromedial prefrontal cortex (mPFC), posterior cingulate cortex (PCC), right superior temporal gyrus (STemp), left inferior temporal gyrus (lTemp) and cerebellum at 24 h compared to baseline ($p < 0.05$ FWE, n1⁄453; x and z represent MNI coordinates). Scatter plots depict link between plasma BDNF levels and RSFC changes after 24h compared to baseline divided by subgroups for (B) mPFC and (C) PCC after ketamine (left side) and placebo (right side) infusion. Data points above the horizontal reference line at zero represent subjects from subgroups with an increase in BDNF levels after ketamine or placebo infusion, whereas data points below the horizontal reference line represent the subgroups with a decrease in BDNF levels. Crosses represent the mean and the standard error of mean (SEM) of the subgroups and the whole group (in black, n1⁄453). Dotted vertical lines represent the borders when RSFC changes become significant for the whole group.

(Woelfer et al., 2019)

The fact that BDNF levels are statistically correlated with a decrease in the connectivity between the PCC and mPFC inside the DMN highlights the similarities existing between Ketamine and psychedelics, provoking similar changes in the DMN, BDNF expression and structural plasticity. Therefore, it would be interesting to assess if this correlation between BDNF levels and RSFC's changes could be observed for psychedelics as well.

Furthermore, Woelfer and colleagues suggest that changes in functional connectivity could be depending on mTOR pathway activation occurring after BDNF binding on Trkb, leading to an improvement in glutamatergic transmission (Li M et al., 2018; Kraguljac et al., 2017).
This pathway is actually the one that has been shown by Ly and colleagues to be shared by both psychedelics (5-HT2A receptor agonists) and Ketamine (NMDA receptor antagonists), providing their plasticity-promoting proprieties.

The results of these two studies shed light on the hypotheses put forward by Vollenweider and Kometer in their review about neurobiological mechanisms involved in psychedelics action (Vollenweider and Kometer, 2010) by specifying the precise pathway shared by Ketamine and psychedelics to promote structural but also functional plasticity. They also bring some preliminary details about its role in the behavioral response associated with these substances.

## From structural and functional plasticity to psychological flexibility?

According to Carhart Harris and Nutt, two types of mechanisms (neurobiological and psychological plasticity) that can be observed in psychedelics' antidepressant response could, in fact, be linked (as previously examined research on ketamine suggests too), and the short and long-term antidepressant action of psychedelics appear to be dependent on this phenomenon (Carhart-Harris and Nutt, 2017; Swift et al .,2017; Belser et al., 2017; Davis et al., 2020).
Carhart-Harris and his team have also developed a brain functioning model to explore this link, called "the entropic brain" model (Carhart-Harris et al., 2014), and applied it to MDD major depressive disorder in their article "Serotonin and brain function, a tale of two receptors" published in 2017. This model is based on brain imaging sessions carried out during and after the psychedelic experience. According to this model, what we observe on RS-fMRI (resting-state fMRI) on the neurobiological level, during and after the infusion of psychedelics, correspond and superpose to the behavioral effects of these substances on depression, and reciprocally.

To introduce this hypothesis, the research team recalls that entropy (a concept that is usually applied in thermodynamics) can be defined as the uncertainty and unpredictability of events in a system (Ben-Naim, 2007), and has both a mechanical and perceptual meaning.

Indeed, depressed patients often present a state of mental rigidity (Holtzheimer and Mayberg, 2011), characterized by anhedonia, pessimistic-oriented cognition, and disconnection from emotions (Berman et al., 2011; Holtzheimer and Mayberg, 2011).
In addition, as described above, low levels of neurotrophic factors have been found in depressed patients, as well as deleterious changes in connectivity (rigid reinforcement in certain resting state networks, particularly within the DMN). All these arguments are in line with a lowered brain entropy, according to the model presented by Carhart-Harris and Nutt.

Conversely, patients who have received a psychedelics infusion express a sense of liberation, the feeling of having been “reset”, reconnecting with their emotions and a better acceptance towards it (Turton et al., 2014; Watts et al., 2017; Roseman et al., 2017b, Bahi 2018, Davis et al., 2020, Swift et al., 2017). These elements are consistent with a more flexible, plastic, emotional state. At the same time, beneficial functional modifications are observed within resting state networks and an increase in neurotrophic factors expression, like BDNF, are also notified: These states correspond to a restored or even increased brain entropy according to the model presented by Carhart-Harris and Nutt.

To further assess this hypothesis, Carhart-Harris’s team set out to measure the variance in functional connectivity patterns in several identified neural networks, during a psychedelics infusion.

Cerebral entropy measurements inspired by the “free energy principle” by Karl Friston (Friston 2010) were then carried in 9 canonical networks, including the DMN, during Psilocybin or placebo intravenous infusion:

The measurement of entropy within neural networks was carried out in a placebo-controlled clinical trial, on 15 healthy volunteers. An intravenous psilocybin or placebo infusion was delivered for 60 seconds, halfway through two 12-minutes bold fMRI brain imaging sessions with eyes closed, on different days.
To estimate this entropy, the researchers first measured the variance in the state of synchrony within different networks: if the signal between all the voxels of a given network deviated very little compared to the global signal, then the variance was considered low within this network; if the signal of the network deviated erratically and significantly, then the variance was considered high within this network.
During the psilocybin infusion, the variance was increased within the networks associated with high levels of cognition, but not in the networks associated specifically with sensory or motor activities. No significant change was found during the placebo sessions.
These results are consistent with the entropic brain model (Carhart-Harris et al., 2014).

Carhart-Harris’ team then sought to express these results as a formal measure of entropy, by calculating the probability distribution of the variance of the intra network synchrony across time: They discretized the time course of intra-network synchrony over time in virtual equal-sized bins and each time point was entered into a bin depending on the variance in the network synchrony at that time point. Then, Shannon’s entropy was calculated for each network (Carhart-Harris et al., 2014).
As expected, results showed that there was a significantly greater entropy in the high-level association networks too, particularly within the DMN (Figure 6, Carhart-Harris et al., 2014).
Moreover, the connectivity patterns of the network during psilocybin infusion were significantly more random and unpredictable than those observed with placebo (Carhart-Harris et al., 2014).

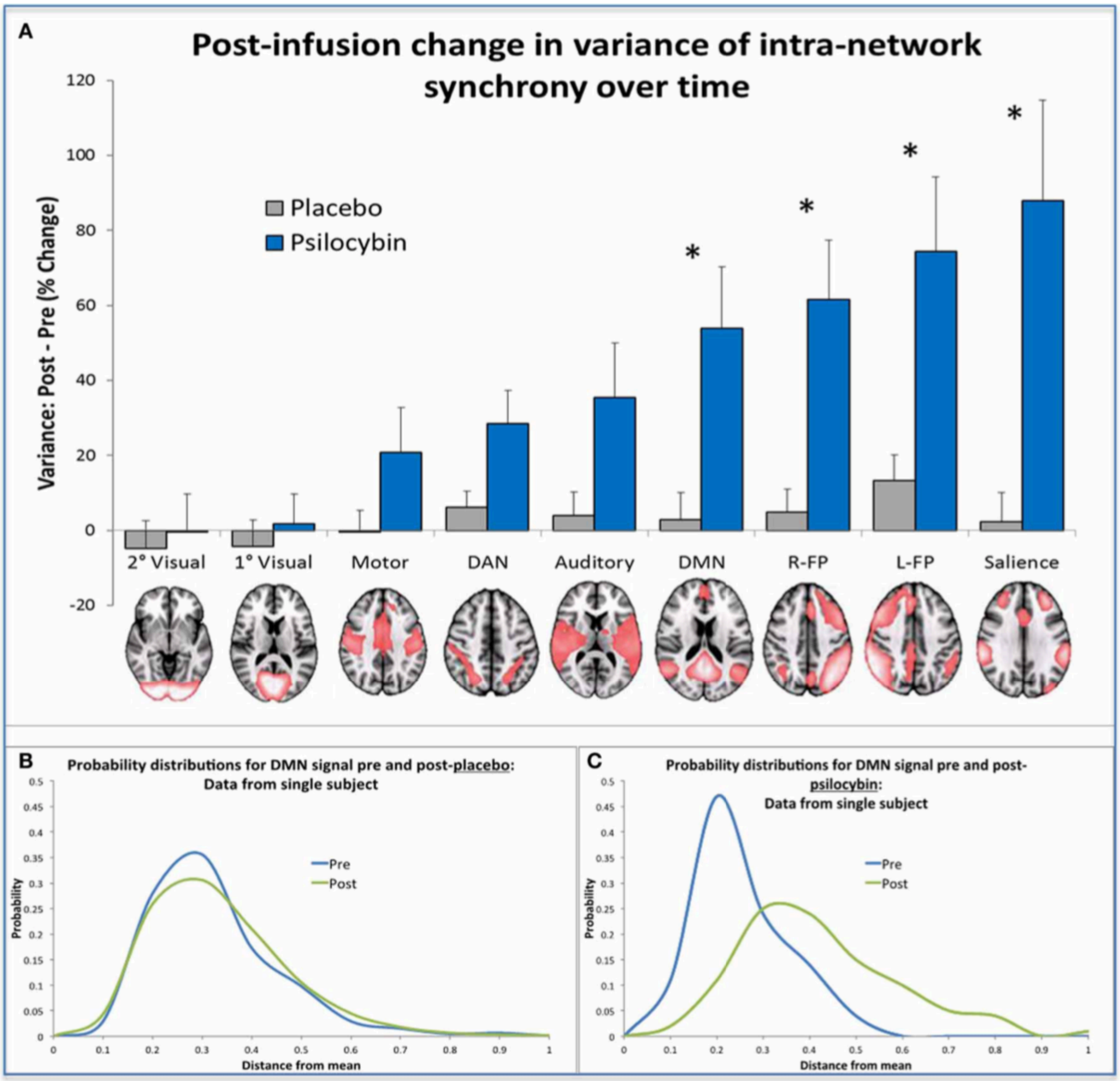

**Figure 6. Changes in network metastability and entropy post-infusion of psilocybin.**

(A) This chart displays the mean variance of the internal synchrony of 9 brain networks for the sample of 15 healthy volunteers, as a percentage change post vs. pre-infusion. A post-infusion increase in metastability for a specific network indicates that the mean signal in that network is a poor model of the activity in its constituent voxels, implying that the network is behaving more “chaotically” post-infusion than pre. Bonferonni correction gave a revised statistical threshold of $p < 0.006$ (0.05/9). One-sample (2-tailed) $t$-tests were performed, comparing the % change against zero. The significant networks are labeled with an asterisk.

(B,C) These probability distributions were derived from data from the same single subject, by discretizing a measure of the internal synchrony of the DMN across time into bins. These bins reflect the distance a data point is from the mean and this gives a probability distribution of the variance of internal synchrony within a network for a given time period (e.g., a 5min period of scanning). The probability distributions shown in Chart B were produced from placebo data where it is clear that prediction of internal network synchrony of the DMN across time is similar before and after infusion (i.e., the blue and green curves). The probability distributions shown in Chart C were derived using psilocybin data and here it is evident that following infusion of psilocybin (i.e., the green curve), prediction of internal network synchrony within the DMN is more difficult compared to pre infusion (the blue curve). When the entropy change was calculated for the group, significantly greater increases in entropy were found in the same networks highlighted in (A) (post-psilocybin vs. pre) vs. (post-placebo vs. pre).

(Carhart-Harris et al, 2014)

When applied to the "entropic brain" model, these results suggest that 5HT2A receptors stimulation could constitute a way to increase brain entropy, through disintegration and reintegration of particular resting-state networks, such as the DMN.
To go further assessing a specific pathway connecting molecular, connectivity and behavioral aspects of the psychedelics' mechanism, and using the results of the studies presented earlier, it may be suggested that this increase in entropy within specific brain networks could be linked to the action of BDNF and mTOR pathways: this process would then promote a more flexible and plastic state, at a neurobiological, but also at a cognitive and behavioral level, thus constituting a key mechanism in psychedelics' effectiveness against MDD.

Therefore, this could allow a "reset" in the psychological mechanisms reinforced during the onset of major depressive disorder that led the patient to this state of rigidity –depicted by a pessimistic vision of himself and his environment (Moutoussis et al., 2014)– and reduced the ability to cope with personal difficulties.
The notion of network reinitialization is also explored through similarities between acute functional connectivity changes observed over 24h with both psilocybin and ECT (Carhart-Harris and Nutt, 2017).
This reset then favors a more malleable and flexible cognition, allowing the patient to benefit more fully from psychological and cognitive therapies, giving him the possibility to get out of his pessimistic perceptions and take advantage of his personal difficulties, receiving the benefits that these life events could provide.

Through this process, mental rigidity is then alleviated, in favor of flexibility and plasticity, due to a mechanism that could be compared, according to the analogy developed by Carhart-Harris, to the one that allows an apparently inflexible metal piece to be shaped by heating it, increasing the entropy within it, allowing it to become malleable (Carhart-Harris and Nutt, 2017).

A study that has been recently published (January 2020) conducted by Davis and colleagues at the Johns Hopkins University also adds weight to this hypothesis.
This study was conducted on 985 volunteers who used psychedelics during their lifetime and was carried out in the format of an online survey, to explore a possible correlation between the acute effects of psychedelics and an increase in psychological flexibility, as well as between an increase in psychological flexibility and a decrease in anxiety and depression scores (Davis et al., 2020).

DASS-21 (depression, anxiety and stress scale) was used to measure the corresponding factors and the Acceptance and Action Questionnaire 2 (AAQ-2) was expressed in positive values to measure the psychological flexibility.
MEQ30 mystical experience scale and Psychological Insight Questionnaire (PIQ; that was created for this purpose) were also used to explore a possible correlation between the increase in psychological flexibility, the occurrence of mystical experience and "psychological insight" (Davis et al., 2020).

Results show a significant correlation between the acute effects of psychedelics, an increase in psychological flexibility, reduction in depression and anxiety scores, and verify the statistical hypotheses according to which:
"acute psychedelic effects (mystical and insight effects) predict a decrease in depression and anxiety following a psychedelic experience" (H1), "acute psychedelic effects predict the increase in psychological flexibility in patients" (H2), "the increase in psychological flexibility predicts a decrease in depression and anxiety scores" (H3) and finally, "the increase in psychological flexibility represents a mediator between Psychedelics acute effects and the decrease in depression and anxiety associated with these substances" (H4) (figure 9) (Davis et al., 2020).
These results need to be verified in laboratory settings and on a more diversified sample of patients, but they already give a concrete psychological appreciation of the hypothesis formulated by Carhart-Harris, by quantifying and assessing the development of psychological flexibility after psychedelics' administration, which is beneficial in reducing depressive and anxious symptoms.

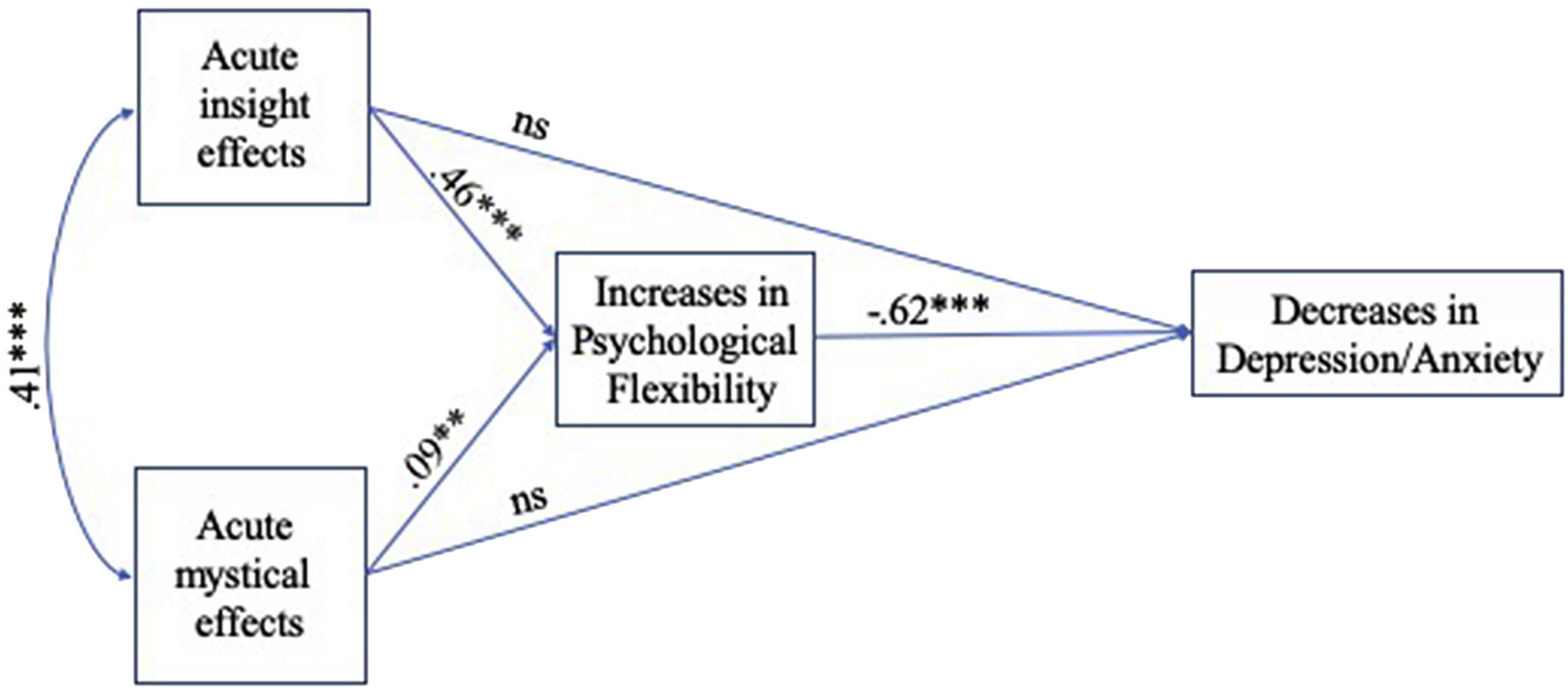

**Figure 9. Path analysis showing that increases in psychological flexibility mediate relationship between acute effects of psychedelics and decreases in depression/anxiety (full sample, N=985). **p<.01, ***p<.001.**

(Davis et al., 2020)

Finally, to further support the global model suggested in this review, it is relevant to specify that Wolff and colleagues proposed a psychological model underlying this increase in psychological flexibility. This model suggests that the psychedelic experience could promote acceptance over aversion based coping methods and involves fear extinction learning as well as beliefs-relaxation (Wolff et al., 2020). This theory is in line with the hypothesis formulated by Ly and colleagues, linking the neurobiological and pharmacological effects of psychedelics to their behavioral outcomes through fear extinction learning, but also add a cognitive explanation to the entropic brain model and the associated psychological flexibility that it involves by linking increased entropy to beliefs-relaxation.

# Conclusion

This review presents several limitations that need to be discussed:
Firstly, it explores an innovative direction regarding what is known and done about major depressive disorder, focusing on signaling pathways that have been little envisaged so far and for which precise mechanisms remain unclear. Moreover, all the results from the articles exposed in this review remain to be reproduced on a larger number of patients, and some need to be confirmed in a human model.
Several studies also require to be reproduced in patients with depression, such as the study conducted by Woelfer and colleagues, exploring the correlation between peripheral BDNF evolution and changes in functional connectivity after Ketamine infusion. It is important to note that, although the peripheral BDNF levels are correlated to the central levels, knowledge about BDNF remains incomplete. This is also the case for the study conducted by Carhart-Harris et al., presenting the cognitive model "the entropic brain" applied to depression.

In addition, it is possible to criticize this review for being reductive in its approach: exploring the role of neuroplasticity in depressive disorder, by connecting the effects of psychedelics on structural and functional plasticity to their psychopharmacological mechanisms (involving 5-HT2A signaling, BDNF –mostly cortical– and mTOR pathway) and to the behavioral and psychological manifestations superposing to these functional changes.
Indeed, this approach –despite holding a central position among the pathophysiological and clinical aspects of depression– does not take into account the full complexity of this pathology and the broad range of mechanisms that it implies.

However, this perspective could be useful in the sense that it gives place to isolate and highlight some mechanisms that can serve both as potential targets in the framework of depression treatment, as a starting point for developing more therapeutic tool against this pathology as well as a way to develop biomarkers in psychiatry (e.g. for antidepressant response). It also gives further details about several hypotheses formulated several years ago, concerning psychedelics' mechanism of action (Vollenweider and Kometer, 2010).
These steps are important for allowing the development of new and effective ways to fight depression, a process that has been slowed down for years.

Another limitation in this review is that it relies on connections established between the molecular, functional and psychological levels, which are based on correlations and similarities between different therapies, but it is too early to establish the existence of a direct causal link.

Indeed, the mechanism put forward by Woelfer and colleagues underlying the increase in peripheral BDNF levels and changes in functional connectivity after Ketamine infusion is still unclear and its possible application to Psychedelics remains to be confirmed, although several arguments support this hypothesis (e.g. the fact that they share the same signaling pathways for promoting plasticity).
The correlation between functional changes and psychological improvements must also be rigorously explored and this task could be complex, from a methodological point of view.

This endeavor could nevertheless reveal crucial information concerning the role of BDNF and mTOR signaling pathway in the behavioral aspects of major depressive disorder.

However, this review highlights the opportunities that the study of 5-HT2AR mediated cerebral plasticity brings, offering important clues about the specific role that these receptors play in MDD. It also reveals important information about the interaction between different neurotransmitter systems (glutamate and serotonin), particularly through mTOR pathway activation via TrkB, due to an increase in cortical BDNF levels occurring after 5-HT2A stimulation by psychedelics. These properties appear to be shared by both 5HT2A receptor agonists and NMDA receptor antagonists.
This pathway promotes structural neuroplasticity, but also seems responsible for functional connectivity changes in resting state neural networks involved in depression.
These functional and structural plasticity changes appear to constitute a neurobiological support that superposes with the behavioral improvements associated with the antidepressant response provided by psychedelics.
Connecting these dots could furthermore provide tracks to clarify the mechanisms of action possibly shared by several non-pharmacological techniques such as neuromodulation (e.g. ECT, tDCS) or even mindfulness meditation techniques.

Through the exploration and connection of these mechanisms, this review partially lifts the veil that obscures the question of the neurobiological support underlying the efficacy of substances that have long been “controversial”, against MDD. It also highlights the advantage of having a transdisciplinary and holistic view on psychiatry and major depressive disorder; bringing pharmacological, cognitive and psychological aspects together.

Furthermore, this review contributes to reinforcing the legitimacy of conducting research on this topic, by presenting a small fraction of the great amount of fundamental and translational research undertaken on psychedelics and Ketamine for mental health as well as underlining the potential offered by these substances to get a better understanding of depression, and, more globally, of general brain function.

In the future, it would be interesting to define to what extent BDNF and mTOR’s signaling pathway modulation (triggered by 5-HT2A-R activation and NMDA antagonism) could play an important role in depression.
To make it possible, it would be pertinent to explore how the decrease in BDNF levels observed in both the HPA and inflammatory hypotheses could be related to the mental rigidity manifested by depressive patients.
It would also be relevant to explore the apparently distinct roles of cortical and hippocampal BDNF in depression and how 5HT2A-R agonism leads to an overexpression of cortical BDNF levels.

Studying the differences between the pathways that are seemingly shared by Psychedelics and Ketamine would also be a great way to develop future useful therapies, understanding, for instance, why Ketamine is addictive while Psychedelics are not, and then being able to target BDNF overexpression and mTOR stimulation using the most beneficial and safe gateway.

Finally, establishing which psychological and cognitive methods should be used to obtain maximized benefits from the psychedelic therapy is also an important concern, and therapeutic methods that aim to promote acceptance seem to be an interesting track to explore (Wolff et al., 2020).

## Acknowledgements

I would like to thank Pr. Emmanuel Haffen, MD, PhD; Dr. Julie MONNIN, PhD and Pr. Virginie NERICH, PharmD, PhD from the University of Franche-Comté for having reviewed my project in its French version, in the framework of my Pharmacy Master thesis. I would also like to thank Susanna PEREZ BLAZQUEZ and Kyle GAMACHE for the help they provided in reviewing the English translation of this paper.
I would like to address special thanks again to Pr. Emmanuel HAFFEN for his trust, letting me work on this fascinating, yet innovative topic.
Finally, I would like to thank Dr. Max WOLFF, PhD and Dr. Henrik JUNGABERLE, PhD for the inspiring conversations they offered around this topic.